\documentstyle[aps,pre,epsf]{revtex}
\begin{document}



\draft
\title{Quasi-classical Molecular Dynamics Simulations of the Electron Gas: Dynamic properties}
\author{
        J. Ortner ,
        F. Schautz, and W. Ebeling} 
\address{Humboldt Universit\"at zu Berlin, Institut f\"ur Physik\\ Invalidenstr.110, D-10115 Berlin}
        
\date{Received 28 May 1997}
\maketitle

\begin{abstract}
Results of quasi-classical molecular dynamics simulations of the quantum electron gas are reported. Quantum effects corresponding to the Pauli and the 
Heisenberg principle are modeled by an effective momentum-dependent
Hamiltonian. The velocity autocorrelation functions and the dynamic 
structure factors have been computed. A comparison with theoretical
predictions was performed.\\

\end{abstract}
 
\section{Introductions}

A gas of interacting electrons imbedded in a uniform positive background 
is a basic theoretical model. Degeneracy (i.e. the influence of 
the Fermi statistics) and coupling (the role of interactions due to Coulomb
forces) are important features of the electron plasma. The electron gas is 
characterized by 2 parameters - the density n and the temperature T (in 
energy units). The dimensionless parameter \begin{math} \theta = \frac{T}{E_F}
\end{math} , \begin{math} E_F = \frac{\hbar^2}{2m}(3\pi ^2 n)^{2/3} \end{math}
being the Fermi energy, describes the degeneracy of the electron system. The
coupling constant of the electron plasma can be defined as the ratio of the 
average Coulomb energy to the average kinetic energy,
\begin{equation} \label{1}
\gamma = \frac{3}{2} \frac{e^2/a}{\varepsilon_{kin}} = \frac{e^2}{a T_{eff}},
\end{equation} 
where
\begin{equation} \label{2}
T_{eff} = \theta^{3/2}T F_{3/2}(\mu /T)
\end{equation}
is the effective temperature corresponding to the kinetic energy of the Fermi gas, $\mu$ is the chemical potential of the ideal electron gas, \begin{math} a= (3/4 \pi n)^{1/3}
\end{math} and \begin{math} F_{3/2} 
\end{math} is the Fermi integral.

\begin{samepage}
For a strongly degenerate electron gas (\begin{math} \theta \ll 1 \end{math}) the
coupling constant depends only on the density,
\begin{equation} \label{3}
\gamma_d = \frac{e^2/a}{\frac{2}{5}E_F} = 1.3575 r_S \,\, ,
\end{equation}
where \begin{math} r_S = a/a_B \end{math} is the Wigner-Seitz radius of the 
electron system in units of the Bohr radius.

The coupling constant for an electron gas obeying classical statistics 
(\begin{math} \theta \gg 1 \end{math}) is
\begin{equation} \label{4}
\gamma_{cl} \equiv \Gamma = \frac{e^2}{aT} \qquad .
\end{equation}
\end{samepage}

Important informations on the properties of the electron gas can be obtained 
from the knowledge of the dielectric function and the dynamic structure factor. The plasma dispersion relation, static correlation functions, and thermodynamic potentials can all be obtained from these functions.

For a weakly coupled plasma (\begin{math} \gamma \ll 1 \end{math}, i.e. for 
very high electron densities or vice versa at very small densities and 
sufficient high temperature) the random phase approximation (RPA)\cite{{Lindhard},{Arista}} gives good 
results in calculating the dielectric function because the kinetic energy is 
dominant.

On the contrary, for electron densities corresponding to the conduction band 
density in real metals the average kinetic and potential energies of the 
electrons are of the same order of magnitude, and the plasma is so to say 
moderate coupled (\begin{math} \gamma \sim 1 \end{math}). The microscopic
dynamics of electrons in such plasmas are dominated by collisions and the 
non-collisional RPA becomes inapplicable. Exchange and correlation corrections 
to the RPA cannot be calculated exactly, but a number of approximate expressions
of the dielectric function have been suggested \cite{Ichimaru}. 

In order to check the validity of the different approaches microscopic 
simulations of the electron gas are of great interest. Classical simulations of the one component plasma were performed by Hansen et al. \cite{Hansen2}.

The aim of this paper is to report the results of a series of quasi-classical molecular 
dynamics simulations  on the dynamic properties of the electron gas. The thermodynamic properties (mean 
energy) of the electron gas were investigated in a previous paper \cite{Ebeling1}.

\section{The model}
In order to treat the quantum electron gas by quasi-classical simulations we make use of effective pair potentials. Purely space-dependent effective
potentials can be derived from the quantum-mechanical Slater sum \cite{{Kelbg},{Deutsch}} . At short 
distances these potentials differ from the bare Coulomb potential and remain 
finite. On the basis of such potentials Norman and Valuev \cite {Norman} and Hansen and 
Mc Donalds \cite{Hansen1} performed molecular dynamics simulations of an electron-proton 
plasma.

Another way to include quantum diffraction effects (i.e. the Heisenberg 
principle) into the dynamics is to blow up the phase space by introducing 
additional "quantum" degrees of freedom. This is done in the wave-packet 
dynamics developed to describe nuclear collisions \cite{{Feldmaier1},{Feldmaier2},{Klakow1},{Klakow2}}. However, this method 
leads to difficulties in describing thermal equilibrium properties of many particle 
systems \cite{Ebeling2}. 

We use therefore in our calculations the ordinary 6N dimensional 
phase space, the particles interacting by effective pair potentials. However, 
a pseudopotential depending only on the space coordinates leads necessarily to 
the Maxwell momentum distribution. To model the momentum distribution of an 
electron gas governed by Fermi statistics we include in our simulations 
momentum-dependent interaction terms. We thus follows a line developed by a 
series of authors as e.g. Wilets and Kirschbaum, Dorso et. al. \cite{{Dorso1},{Dorso2},{Kirschbaum},{Wilets}}.

In our simulations we substitute the quantum dynamics of the electron system by a phase space dynamics of Hamilton type with effective quasi 
classical Hamiltonian \cite{Ebeling1}
\begin {equation} \label {5}
H \, = \, \sum_{i=1}^N \frac {p_i^2}{2m} \,\, 
       + \,\,\sum_{i < j}^N \, V_p(\frac{r_{ij}}{r_0}, \frac{p_{ij}}{p_0}) \,\,
       + \,\, \sum_{i < j}^N \, e^2 \, F(\frac{r_{ij}}{r_0}, \, \frac{p_{ij}}{p_0})
\qquad .
\end {equation}
Here the first term is the ordinary (classical) kinetic energy of the electrons. 
The second contribution, the Pauli potential was chosen in a form suggested by
Dorso et al. \cite{Dorso1},
\begin {equation} \label {6}
V_p(p,r)\,=\, V_0 e^{-\Delta^2/2}
\qquad ,
\end {equation}
where $\Delta^2 \,=\, \frac {p^2}{p_0^2} \,+\, \frac {r^2}{r_0^2} $ is the 
effective phase space distance of two particles with relative momentum p and 
distance r.

The last term in the effective Hamiltonian is the Coulomb interaction averaged 
with respect to the two particle Gaussian wave packets and is 
expressed by
\begin {equation} \label {7}
F(r,p) \,=\, \frac {erf(r/ \sqrt{2} \, r_0)}{r} 
\qquad .
\end {equation}
The Gaussian wave packet transforms the Heisenberg uncertainty condition into
an identity, $(\delta p)(\delta q) = \hbar / 2$ which leads to
\begin {equation} \label {8}
r_0 p_0 \, \, = \, \, \hbar 
\qquad .
\end {equation}
The other two parameters in the Hamiltonian Eq. (5) are chosen to describe 
the correlation function and the momentum distribution of a free Fermi gas. An appropriate choice is \cite{Ebeling1}
\begin {equation} \label {9}
V_0=T_{eff} \qquad , \qquad p_0^2 \,=\, m T_{eff} \qquad , \qquad
r_0^2 \, = \, \frac{\hbar^2}{mT_{eff}} \qquad ,
\end {equation}
with $ T_{eff} $ from Eq. (2),

The simulations based on the Hamiltonian 
Eq. (5) with the parameters defined by Eq. (\ref{9}) result in a mean energy proved 
to be in good agreement with Quantum Monte Carlo simulations and with Pade 
approximations \cite{Ebeling1}. 

That is the reason why we expand our considerations to the 
investigation of the dynamic properties of the electron gas within the 
developed approach.

In restricting our calculations to a simple Hamiltonian given by Eq. (5) we 
make use from the fact that the collective dynamics of the electron system 
are dictated primiraly by the long range character of the Coulomb potential and 
are widely unaffected by the simplifications in the short range part of the 
effective potentials made in Eqs. (6) and (7).

\section{Dynamic properties}

Let
\begin{equation} \label{10}
\rho(\vec{k},t) \, = \, \sum_{i=1}^N \, \exp (i \vec{k} \cdot \vec{r_i} (t))
\end{equation}
be the Fourier component of the time-dependent microscopic electron density. The 
density-density dynamic structure factor is defined as the Fourier transform of 
the correlation function,
\begin{equation} \label{11}
S(\vec{k} ,\omega) \, = \, \frac{1}{2 \pi N} \int_{- \infty}^{\infty} \, 
e^{i \omega t} \, <\rho(\vec{k} ,t) \, \rho(- \vec{k} , 0)> \, dt \qquad .
\end{equation}

A closely related quality is the dielectric function 
$ \varepsilon (\vec{k}, \omega) $ of the electron system. It describes the 
linear response of the plasma to an external electric field and is connected to 
the dynamic structure factor via the fluctuation - dissipation theorem (FDT),
\begin{equation} \label{12}
S(\vec{k}, \omega) \, = \, \frac{\hbar Im \, \varepsilon ^{-1} (\vec{k}, \omega)}
{n \pi \phi (k) \,  [1 \, - \, \exp (- \beta \hbar \omega)]} \qquad ,
\end{equation}
where $ \phi (k) = 4 \pi e^2 / k^2 \qquad , \qquad \beta = 1/T $ .

The imaginary part of the dielectric function is antysymmetric with respect to 
the frequency. Note that from Eq. (12) it follows that the dynamic structure 
factor does not possess symmetry, but satisfies the relation
\begin{equation} \label{13}
S(\vec{k}, - \omega) = e^{- \beta \hbar \omega} \, S(\vec{k} , \omega) \qquad .
\end{equation}
The dynamic structure factor defined by Eqs. (11) and (10) is directly 
measurable in the MD simulations if one identifies the Heisenberg operator 
$ \vec{r_i} (t) $ with the position of the i-th particle in our simulations.

However, the thus obtained quantity (we denote it by $ R( \vec{k} , \omega) $) possesses symmetry.
It corresponds therefore to a classical FDT,
\begin{equation} \label{14}
R (\vec{k} , \omega) = (n \pi \phi (k) \beta \omega)^{-1} \, 
Im \, \varepsilon^{-1} (\vec{k}, \omega) \qquad .
\end{equation}

From Eq. (\ref{14}) one concludes that $R (\vec{k} , \omega)$ can be regarded as a normalized loss function. 

The quantum-statistical dynamic structure factor obeying the relation Eq.(13) has to be 
calculated as
\begin{equation} \label{15}
S (\vec{k} , \omega) = \frac{\hbar \beta \omega}{1 - exp ( - \beta \hbar \omega)} \, 
R(\vec{k} , \omega) \qquad .
\end{equation}\\

In what follows we will regard the normalized loss function. Note, that in the classical case the loss function and the dynamic structure factor coincide.

In our molecular dynamic simulations we used the algorithm of Verlet \cite{Verlet} to integrate numerically the equations of motions obtained from the effective Hamiltonian of a system of 256 electrons. The typical length of the MD runs were about $10^3 \omega_p^{-1}$ ($\omega_p$ being the plasma frequency). 

The equilibrization phase was replaced by a Monte Carlo Simulation using the algorithm of Metropolis \cite{Metropolis}.

The forces were calculated by an Ewald method in order to account for the long range of the Coulomb interaction \cite{Brush}.

The motion of the electrons can be studied by calculating the velocity autocorrelation function $<v(t+\tau) v(t) >_t$. 

We see (Fig.1) that for $\Gamma=1$ the velocity autocorrelation falls monotonically to zero , whereas for $\Gamma=100$ the decay of the velocity ACF is characterized by appearance of oscillations with a frequency close to the plasma frequency. That means that the motion of a single electron is coupled to the collective density fluctuations.\\
The collective motion is described by the dynamic structure factor (or the loss function).
$ R(q,\omega) $ ($ q=ka $) is plotted for two q values, at $ \Gamma=1 $ for 
different $ \theta=1 $ (moderate degenerate) and $ \theta = 50 $ (classical)
and for $ \Gamma = 100 $ at $ \theta = 50 $ 
(strongly coupled, classical electron gas)(Figs.2-4). The results of the simulations are 
compared with theoretical predictions from RPA.  At moderate coupling constants ($ \Gamma = 1 $) the plasmon 
peak of the loss function is less than that predicted by RPA and slightly 
shifted to the left(Figs.2,3). In both cases the plasmon peak can be observed 
only for the smallest q value. The change of $ \theta $ in the range from 50 to 1 has only 
small influence on the results (Fig.5). 

In the strong coupling regime $ \Gamma = 100 $ 
the observed plasmon peak at the smallest q value is extremly sharp and centered 
close to $ \omega_P $ (Fig.4). A well defined collective plasmon mode has been 
developed. At a q value 3 times larger the plasmon peak widens, but is still 
present and shifted to the left by about 10 per cent. At still larger q values 
the plasmon peak vanishes. 

This behavior is in striking contradiction to the 
RPA predictions where no plasmon peak can be observed due to the strong Landau 
damping. However, the RPA is inapplicable in the strong coupling regime, where
the potential energy is dominant.  On the contrary, our MD simulations for the case of weak degeneracy are in a 
good agreement with corresponding MD simulations of Hansen et al. for the classical one component plasma \cite{Hansen2}.

We conclude therefore that our model yields reasonable results in describing the dynamic properties of the electron gas at least at weak and moderate degeneracy.

\section{Analysis of the results}
Important characteristics are the frequency moments of the imaginary part of 
the inverse dielectric function(DF). They are defined by
\begin{equation} \label{16}
C_{\nu}(\vec{k}) = - \frac{1}{\pi} \, \int_{- \infty}^{\infty} \, \omega ^{\nu - 1}
\, Im \, \varepsilon ^{-1} (\vec{k} , \omega) \, d \omega \qquad ,
\qquad \nu=0,1,\ldots \qquad .
\end{equation}

Due to the antisymmetry of the imaginary part of the inverse DF all even frequency moments vanish, 
whereas the odd frequency moments are purely expressable in terms of the static properties 
of the electron gas. After a straightforward calculation one obtains,
\begin{equation} \label{17}
C_0 (k) =  (1 - \varepsilon^{-1}(\vec{k},0)) \qquad ,
\end{equation}
\begin{equation} \label{18}
C_2 (k) =  \omega^2_p \qquad ,
\end{equation}
\begin{equation} \label{19}
C_4 (k) = \omega_p^4 \, (1 + K(k) + L(k)) \qquad ,
\end{equation}
where
\begin{equation} \label{20}
K(k) = 3(k/k_D)^2 + {\sqrt{\pi/18} (\lambda_T^3 k^2/\lambda_L)} + 
{\lambda_T^2 k^4 k_D^{-2}} 
\end{equation}
is the kinetic  contribution involving quantum corrections,
$ k_D^2 = 4 \pi n e^2 \beta $ , $ \lambda_T = ( \hbar \beta / 2 m)^{-1/2} $ 
, $ \lambda_L = 3/2 e^2 \beta $ and $ \omega_p^2 = 4 \pi n e^2 / m $ .
The contribution 
\begin{equation} \label{21}
L(k) = \frac{1}{3 \pi^2 n} \, \int_0^{\infty} \, p^2 \, [S(p) - 1] \, f(p,k) \,
dp
\end{equation}
takes into account the electronic correlations,
\begin{equation} \label{22}
f(p,k) = \frac{5}{8} - \frac{3 p^2}{8 k^2} + \frac{3 (k^2 - p^2)^2}{16 p k^3}
\, \ln \left(\frac{p+k}{p-k} \right) \qquad .
\end{equation}

A quantitative analysis of the results of the MD simulations should be based 
on the frequency moments defined in Eqs.(\ref{17} -\ref{19}).

The Nevanlinna formula of the classical theory of moments constructs a dielectric function which satisfies the known sum rules $ C_0 $ to $ C_4 $ \cite{Adamyan}:
\begin{equation} \label{23}
\varepsilon^{-1} (\vec{ k} , z) \, = \, 1 + \frac{\omega ^2_p (z+Q)}
{z (z^2 - \omega^2_2) + Q (z^2 - \omega^2_1)} \qquad .
\end{equation}
Here $ Q = Q(\vec{k},z) $ is an arbitrary function being analytic in the upper complex half-plane $ Im \, z > 0 $ ($\omega = Re \, z$) and possessing there a positive imaginary part, it also should satisfy the limiting condition: $ (q(\vec{k},z)/z) \to 0 $ , as $ z \to \infty $ within the sector $ \vartheta < arg(z) < \pi - \vartheta $ $ (0< \vartheta< \pi) $ . 

The frequencies $ \omega_1(\vec{k}) $ and $ \omega_2(\vec{k}) $ are defined as respective ratios of the moments $ C_n(\vec{k}) $:
\begin{equation} \label{24} 
\omega^2_1 = C_2/C_0 = \omega^2_p (1 - \varepsilon^{-1}(\vec{k},0))^{-1} \qquad, 
\end{equation}

\begin{equation} \label{25}
\omega^2_2 = C_4/C_2 = \omega^2_p (1 + K(k) + L(k)) \qquad.
\end{equation}

There is no phenomenological basis for the choice of an unique $ Q(\vec{k},z)$ , which would provide an exact expression for $ \varepsilon^{-1}(\vec{k},\omega) $ .

If one is interested in the investigation of the dispersion relation only it sufficies to neglect $ Q(\vec{k},\omega) $, since the damping is small in strongly coupled plasmas. If one puts $ Q(\vec{k},\omega) = 0 $ one obtains the expression of the inverse dielectric function obtained within the quasilocalized charge approach of Kalman \cite{Kalman}. The disadvantage of this choice of Q is that damping is not taken into account. As a result the shape of the dynamic structure factor within this approach is reduced to a simple $\delta$ function peak at the frequency $\omega_2(k)$ . Thus only the peak position but not the shape of the dynamic structure factor can be described appropriately.

The easiest way to go beyond the simple approximation $Q=0$ is to put the function $ Q(\vec{k},\omega) $ equal to its static value:
\begin{equation} \label{26}
Q(\vec{k},z) = Q(\vec{k},0) = i h(\vec{k}) \qquad ,
\end{equation}
where $ h(\vec{k}) $ is connected to the static value of the dynamic structure factor $ S(\vec{k},0) $:
\begin{equation} \label{27}
h(k) = \frac{k^2}{k_D^2} \frac{C_0(k)}{S(\vec{k},0)} \left[ (\omega_2/\omega_1)^2 - 1 \right] \qquad .
\end{equation} \\

From the Nevanlinna formula the loss function reads
\begin{equation} \label{29}
R(\vec{k},\omega) = S(\vec{k},0) \frac{\omega h^2(\vec{k}) \omega^4_1}{\left[ \omega^2 (\omega^2 - \omega^2_2)^2 + h^2(\vec{k}) (\omega^2 - \omega^2_1)^2 \right] } \qquad .
\end{equation} 

Formula Eq. (\ref{29}) interpolates between the low frequency behavior (described by $C_0(k)$ and $S(k,0)$) and the high frequency behavior (given by the moments $C_2(k)$ and  $C_4(k)$) of the loss function. Though there is no direct justification that the above interpolation formula describes the loss function in an appropriate manner also for intermediate frequencies, we expect that Eq. (\ref{29}) reproduces the whole shape of the loss function at least qualitatively.

The analysis of the MD calculations is based on the formula Eq. (\ref{29}). 
To calculate the frequencies $h(q)$ , $\omega_1 (q)$ and $\omega_2(q)$ ($q=ka$) the static structure factors from HNC equations were used. Since the latter are classical equations we have restricted our comparison of the sum rule approach to the MD calculations to the case of weak degeneracy. (In this case the dynamic structure factor $S(q,\omega)$ coincides with the loss function $R(q,\omega)$).

The results of the comparison of the loss function calculated from the MD simulations and that obtained from the Nevanlinna formula are shown for different q vectors at $\Gamma=100$ and $\Gamma=1$, respectively. The agreement between theory and simulations is quite good. The theoretical curves reproduce rather well the variation of the shape of the dynamic structure factor and describe the plasmon peak position in a good manner. However, the agreement of the height of the peaks is less satisfactory (Figs.6-8).We believe that one of the reasons for this disagreement between the results of simulations with theoretical predictions might be the normalization to $S(q,0)$ which is a value rather bad measured in the simulations due to the poor statistics at long times.    

\section{Conclusions}
The results of quasiclassical molecular dynamics simulations of the electron gas using momentum dependent effective potentials have been reported. The quasiclassical MD computations were performed for different coupling constants ($ \Gamma = 1$ and $ \Gamma = 100 $) at various degeneracy ($ \theta=1 $ and $ \theta = 50 $).
The effective potential was chosen to describe both the Pauli principle and the Heisenberg uncertainty.
The changing of the coupling constant $ \Gamma $ leads to qualitative changes in the dispersion curve of the density fluctuations.

At moderate coupling ($ \Gamma = 1$) the dispersion is positive, the plasmon peak is observed only for the smallest wavenumber $ q=0.618 $ . As can be seen from Figs. 2 and 3 this behavior is qualitatively confirmed by the RPA calculations, however, quantitative deviations from the RPA predictions as a slight shift of the plasmon peak to the left were observed.   

In the strong coupling regime the shape of the loss function changes qualitatively (Fig.4). At $ \Gamma = 100$ the dispersion is negative, a very sharp plasmon peak is observed for the smallest wavenumber, with increasing q the peak widens but is present up to $ q=3.1 $. At still greater q the plasmon peak vanishes. 

Thus, the collective behaviour of the electron gas at weak and moderate coupling can be understood as oscillations of the total charge in the Debye sphere, the individual particles in the sphere moving almost independently. This is the regime of the collisionless plasma described by the RPA. On the other hand one can interpretate the collective motion in the strong coupling regime as solid-like collective oscillations when the motion of each of the particles is coupled to the collective oscillations. This is also confirmed by the shape of the velocity autocorrelation function (Fig.1).

In contrast to the qualitative change of the shape of the loss function by varying the coupling constant the variation of $\theta$ in the range from $\theta=50$ to $\theta=1$ has only small influence on the behavior of the loss function (Fig.5).However, we expect a greater influence at higher degrees of degeneracy.

As demonstrated in Figs.6-8 the above features are rather well produced by a simple sum rule analysis.

Finally we note that our quantum molecular dynamic simulations describe  the dynamic properties of the electron gas only approximately.

\section{Acknowledgments}

This work was supported by the DFG (Germany).

The authors acknowledge valuable discussions with G.E. Norman, D. Kremp, V. Podlipchuk, A. Valuev and C. Toepffer.

\begin {figure} [h]
  \begin{center}
    \leavevmode
    \epsfxsize=130mm
    \epsffile{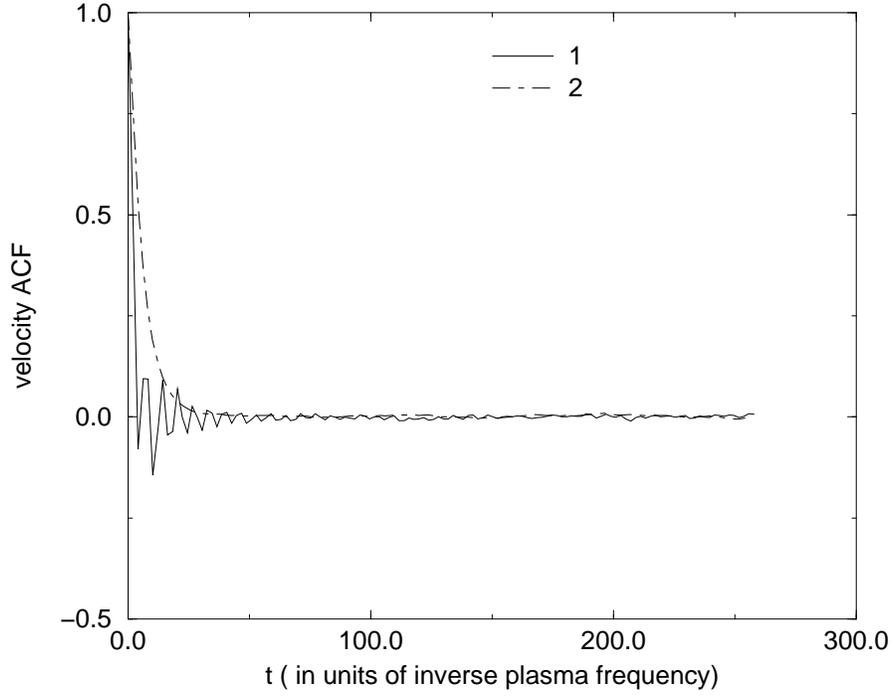}
   \end{center}
\vspace*{-0.7cm}
\caption{velocity autocorrelation function for $\theta=50$ at different$\Gamma$ (1-$\Gamma=100$, \, 2-$\Gamma=1$)}
\end{figure} 

\begin {figure} [h]
  \begin{center}
    \leavevmode
    \epsfxsize=130mm
    \epsffile{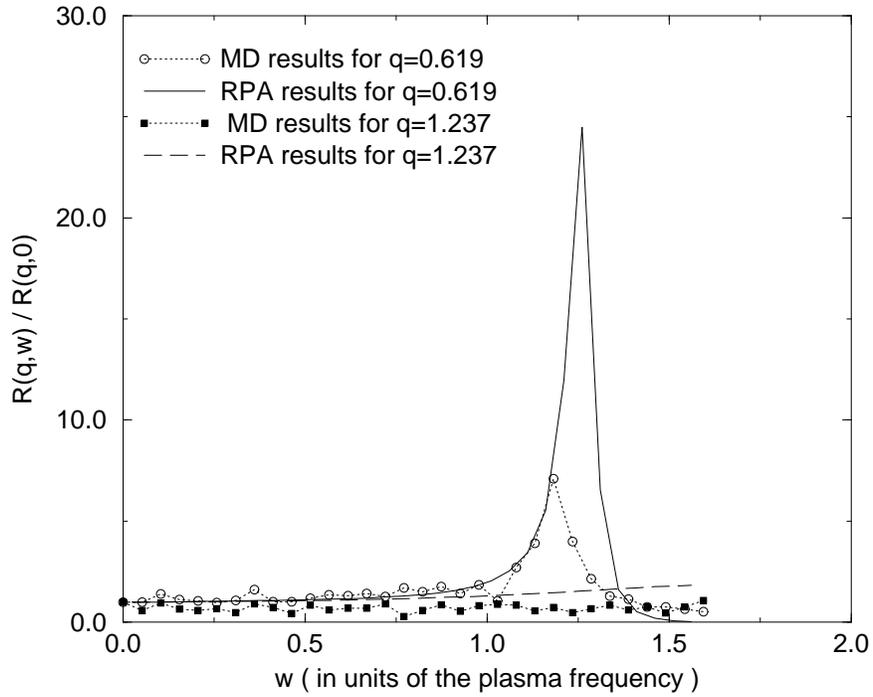}
   \end{center}
\vspace*{-0.7cm}
\caption{comparison of the MD loss function $R(q,\omega)$ versus frequency $\omega/\omega_p$ with the corresponding loss function from the RPA for different wavevectors q at $\Gamma=1$ and $\theta=1$}
\end{figure}

\begin {figure} [h]
  \begin{center}
    \leavevmode
    \epsfxsize=130mm
    \epsffile{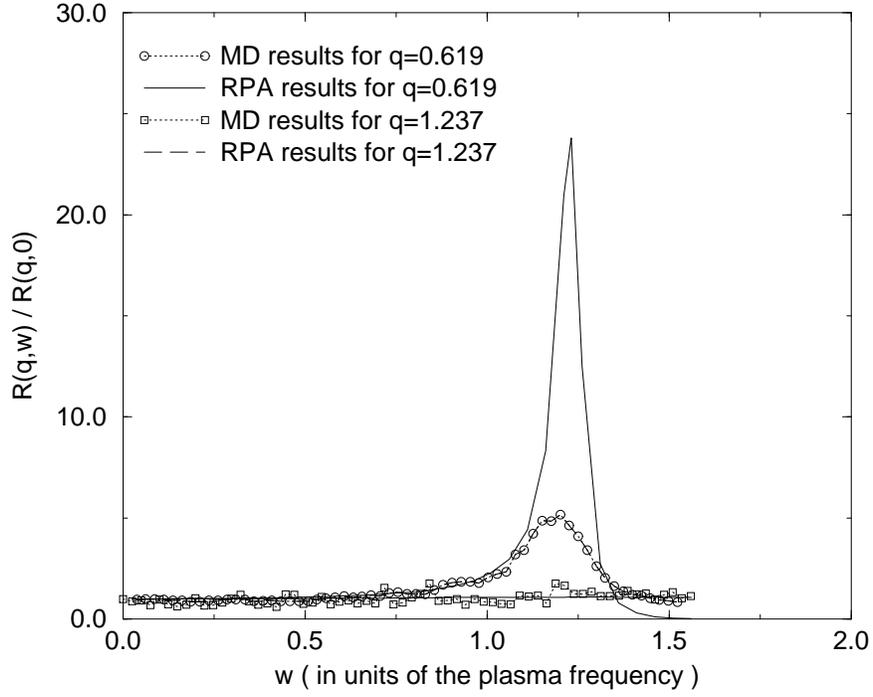}
   \end{center}
\caption{same as Fig.2; for $\Gamma=1$ and $\theta=50$ }
\end{figure} 

\begin {figure} [h]
  \begin{center}
    \leavevmode
    \epsfxsize=130mm
    \epsffile{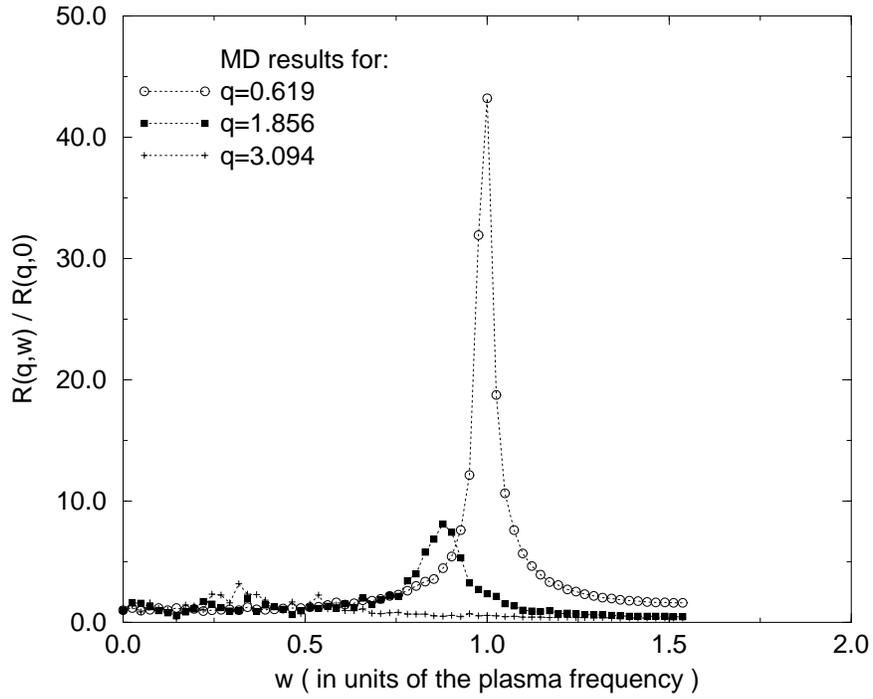}
   \end{center}
\vspace*{-0.7cm}
\caption{The MD loss function $R(q,\omega)$ versus frequency $\omega/\omega_p$  for different wavevectors q at $\Gamma=100$ and $\theta=50$}
\end{figure}

\begin {figure} [h]
  \begin{center}
    \leavevmode
    \epsfxsize=130mm
    \epsffile{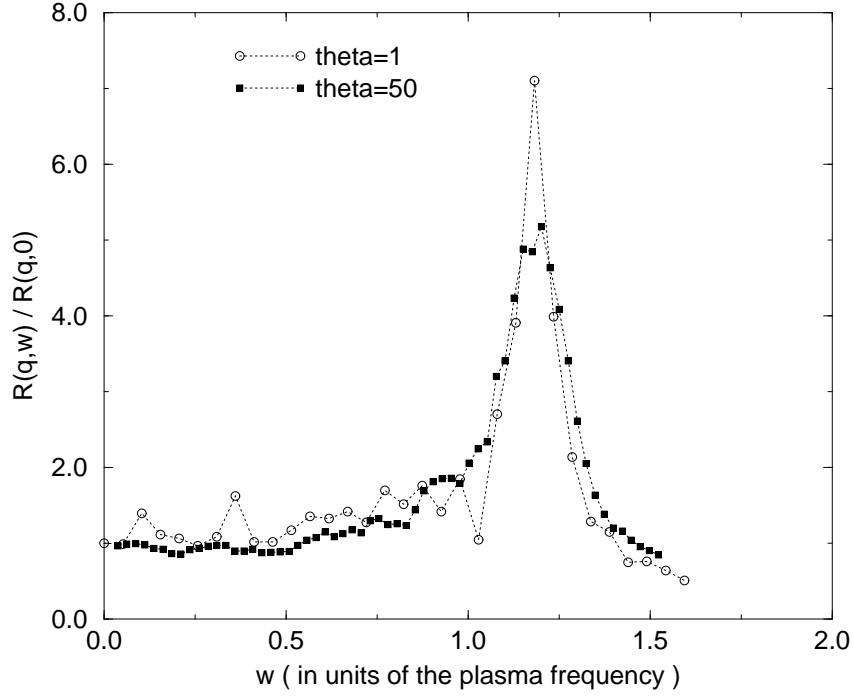}
   \end{center}
\vspace*{-0.7cm}
\caption{The MD loss function $R(q,\omega)$ versus frequency $\omega/\omega_p$  for wavevector $q=0.619$ at fixed $\Gamma=1$ and different $\theta$}
\end{figure}

\begin {figure} [h]
  \begin{center}
    \leavevmode
    \epsfxsize=130mm
    \epsffile{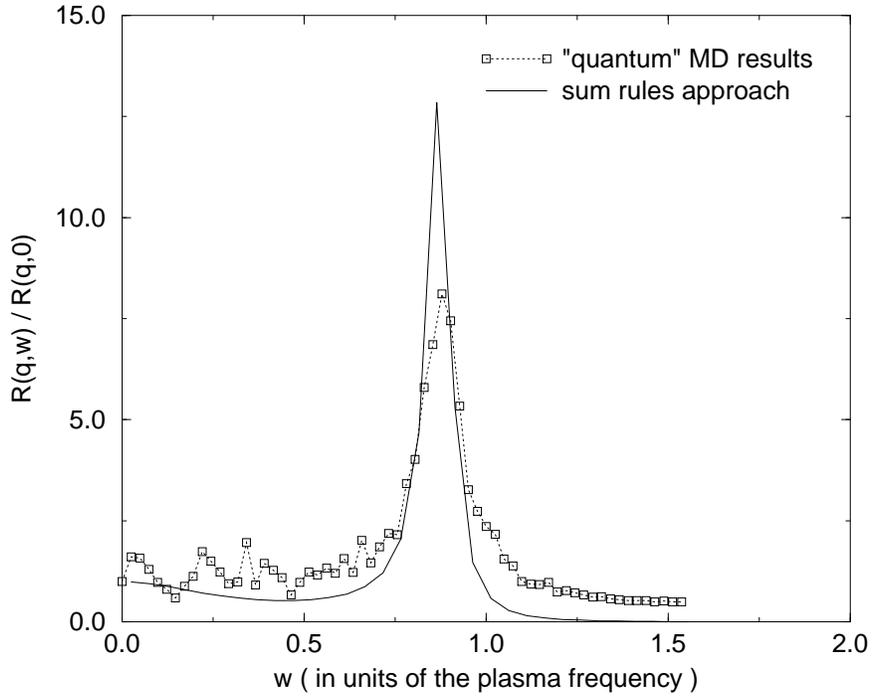}
   \end{center}
\vspace*{-0.7cm}
\caption{comparison of the MD loss function $R(q,\omega)$ versus frequency $\omega/\omega_p$ with the corresponding loss function from the sum rules approach (Eq.29) at $\Gamma=100$ and $\theta=50$ for wavevector $q=1.856$}
\end{figure}

\begin {figure} [h]
  \begin{center}
    \leavevmode
    \epsfxsize=130mm
    \epsffile{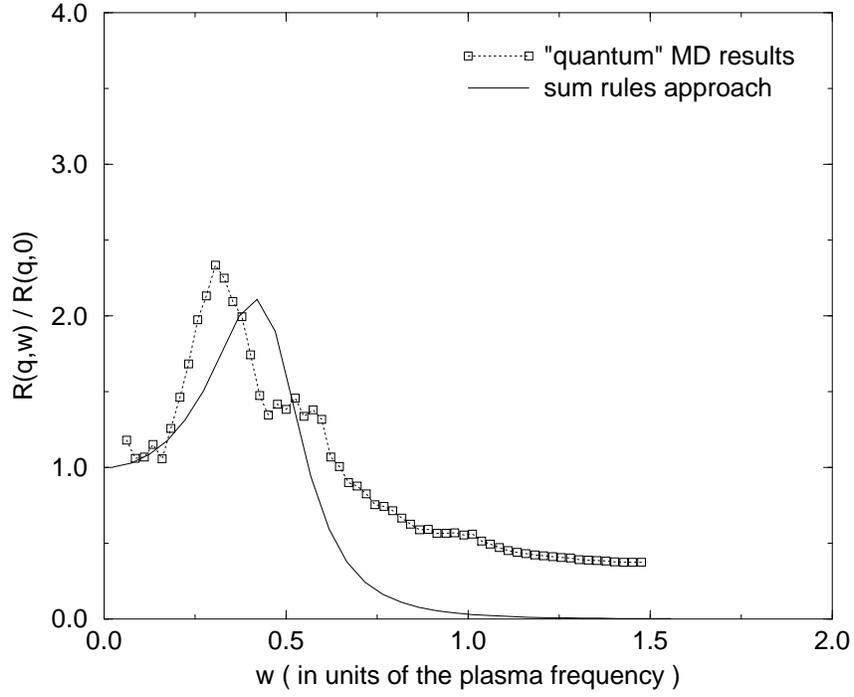}
   \end{center}
\vspace*{-0.7cm}
\caption{same as Fig.6; at $\Gamma=100$ and $\theta=50$ for wavevector $q=3.094$}
\end{figure}

\begin {figure} [h]
  \begin{center}
    \leavevmode
    \epsfxsize=130mm
    \epsffile{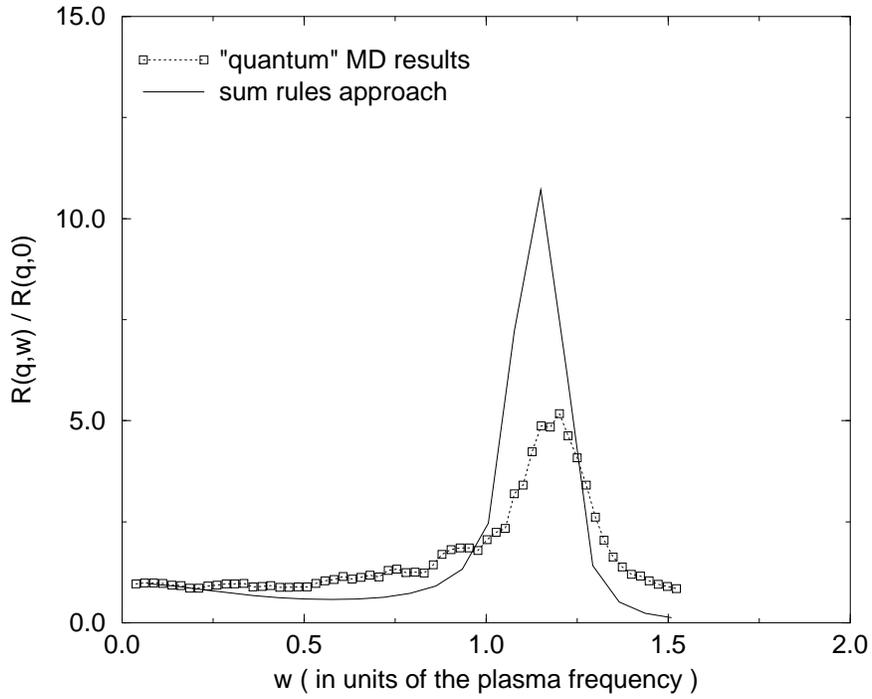}
   \end{center}
\vspace*{-0.7cm}
\caption{same as Fig.6; at $\Gamma=1$ and $\theta=50$ for wavevector $q=0.619$}
\end{figure}

\newpage

List of figure captions
\\
\\
Fig.1. Velocity autocorrelation function for $\theta=50$ at different $\Gamma$ (1-$\Gamma=100$, \, 2-$\Gamma=1$).
\\
\\
Fig.2. Comparison of the MD loss function $R(q,\omega)$ versus frequency $\omega/\omega_p$ with the corresponding loss function from the RPA for different wavevectors q at $\Gamma=1$ and $\theta=1$.
\\
\\
Fig.3. Same as Fig.2; for $\Gamma=1$ and $\theta=50$.
\\
\\
Fig.4. The MD loss function $R(q,\omega)$ versus frequency $\omega/\omega_p$  for different wavevectors q at $\Gamma=100$ and $\theta=50$.
\\
\\
Fig.5. The MD loss function $R(q,\omega)$ versus frequency $\omega/\omega_p$  for wavevector $q=0.619$ at fixed $\Gamma=1$ and different $\theta$.
\\
\\
Fig.6. Comparison of the MD loss function $R(q,\omega)$ versus frequency $\omega/\omega_p$ with the corresponding loss function from the sum rules approach (Eq.29) at $\Gamma=100$ and $\theta=50$ for wavevector $q=1.856$.
\\
\\
Fig.7. Same as Fig.6; at $\Gamma=100$ and $\theta=50$ for wavevector $q=3.094$.
\\
\\
Fig.8. Same as Fig.6; at $\Gamma=1$ and $\theta=50$ for wavevector $q=0.619$.
\end{document}